\documentclass[11pt]{article}

\usepackage{graphics}
\usepackage{amsmath}
\usepackage{authblk}
	
\usepackage[a4paper, total={5.5in, 9in}]{geometry}

\title{Relative distance between tracers as a measure of diffusivity within moving aggregates}

\author[1]{Wolfram P\"onisch}
\author[1,2]{Vasily Zaburdaev}
\affil[1]{Max Planck Institute for the Physics of Complex Systems, N\"othnitzer Stra{\ss}e 38, 01187 Dresden, Germany}
\affil[2]{Institute of Supercomputing Technologies, Lobachevsky State University of Nizhny Novgorod, Gagarina Av. 23, Nizhny Novgorod, 603140, Russia}

\date{}
\begin{document}

\newcommand{\xzeroi}{x_\mathrm{i,0}}
\newcommand{\yzeroi}{y_\mathrm{i,0}}
\newcommand{\dab}{d_\mathrm{ab}}
\newcommand{\dabvec}{\mathbf{d}_\mathrm{ab}}
\newcommand{\ra}{\mathbf{r}_\mathrm{a}}
\newcommand{\rb}{\mathbf{r}_\mathrm{b}}
\newcommand{\razero}{\mathbf{r}_\mathrm{0,a}}
\newcommand{\rbzero}{\mathbf{r}_\mathrm{0,b}}
\newcommand{\ri}{\mathbf{r}_\mathrm{i}}
\newcommand{\dt}{\tau}
\newcommand{\pI}{p_\mathrm{i}}
\newcommand{\pA}{p_\mathrm{a}}
\newcommand{\pB}{p_\mathrm{b}}
\newcommand{\DA}{D_\mathrm{a}}
\newcommand{\DB}{D_\mathrm{b}}
\newcommand{\Di}{D_\mathrm{i}}
\newcommand{\DC}{D_\mathrm{c}}
\newcommand{\xA}{x_\mathrm{a}}
\newcommand{\xB}{x_\mathrm{b}}
\newcommand{\yA}{y_\mathrm{a}}
\newcommand{\yB}{y_\mathrm{b}}
\newcommand{\dvecAB}{\mathbf{d}_\mathrm{ab}}
\newcommand{\xAB}{x_\mathrm{ab}}
\newcommand{\yAB}{y_\mathrm{ab}}
\newcommand{\pAB}{p_\mathrm{ab}}
\newcommand{\DAB}{D_\mathrm{ab}}
\newcommand{\MSRD}{\delta}
\newcommand{\MSRDtime}{\delta_\mathrm{t}}
\newcommand{\pGRAY}{p_\mathrm{rice}}
\newcommand{\pRICE}{p_\mathrm{rice}}
\newcommand{\pRICEsum}{\tilde{p}_\mathrm{rice}}
\newcommand{\pCIRC}{p_\mathrm{circ}}
\newcommand{\pCIRCi}{p_\mathrm{i,circ}}
\newcommand{\pCIRCa}{p_\mathrm{a,circ}}
\newcommand{\pCIRCb}{p_\mathrm{b,circ}}
\newcommand{\pCIRCab}{p_\mathrm{ab,circ}}
\newcommand{\pRAYLEIGH}{p_\mathrm{ray}}
\newcommand{\pGRAYsum}{\tilde{p}_\mathrm{rice}}
\newcommand{\MSRDsaturt}{\delta_\mathrm{sat,t}}
\newcommand{\MSRDsature}{\delta_\mathrm{sat,e}}

\maketitle

\begin{abstract}
Tracking of particles, be it a passive tracer or an actively moving bacterium in the growing bacterial colony, is a powerful technique to probe the physical properties of the environment of the particles. One of the most common measures of particle motion driven by fluctuations and random forces is its diffusivity, which is routinely obtained by measuring the mean squared displacement of the particles. However, often the tracer particles may be moving in a domain or an aggregate which itself experiences some regular or random motion and thus masks the diffusivity of tracers. Here we provide a method for assessing the diffusivity of tracer particles within mobile aggregates by measuring the so-called mean squared relative distance (MSRD) between two tracers. We provide analytical expressions for both the ensemble and time averaged MSRD allowing for  direct identification of diffusivities from experimental data.
\end{abstract}

\section{Introduction}
\label{sec:intro}

In most cases, living matter is organized in the form of multicellular aggregates, agglomerates consisting of many individual cells. Examples range from microcolonies formed by bacteria~\cite{ponisch20172,taktikos2015} (see figure~\ref{fig1}a) to eukaryotic cells forming aggregates~\cite{douezan2011, pampaloni2013, montel2012} and tissues~\cite{merkel2016}.

One of the standard ways to experimentally assess the mechanical properties of such agglomerates is by performing particle tracking and analyzing the trajectories of individual cells or embedded passive tracer particles. Similar measurements can be performed on a subcellular level with injected particles or tracing cell organelles as means of quantifiying the physical properties of the cell cytoplasm~\cite{munder2016ph,tolic2004anomalous}. By assuming a random motion of cells within agglomerates or tracers in the cell cytoplasm one typically measures the mean squared displacement (MSD) and thus gets access to the diffusion constant and the scaling of diffusion. 

However, frequently cell aggregates or individual cells exhibit spatial translation and rotation~\cite{ponisch20172,taktikos2015}. This motion contributes to the MSD of tracers and makes it difficult to disentangle the diffusivity of the tracers.

In this paper we investigate a quantity that enables us to measure the diffusion coefficient of tracers within mobile domains, the so called mean squared relative distance (MSRD). It is similar to the standard MSD except it utilizes the relative distance between two particles. This results in the MSRD, unlike the MSD, being insensitive to the translation and rotation of the domain in which the tracking is happening.
The problem of relative diffusion is more than a century old. From classical works of Richardson and Batchelor ~\cite{richardson1926, batchelor1952, sullivan1971}, to direct applications in biophysical tracking \cite{qian:1991}, this topic is extensively studied. The prototypical quantity of interest is a vector of relative displacement of two tracers. However, the second moment of the displacement carries information about the initial positions of the particles. Normally, when, for example, measuring  two tracers in a turbulent atmosphere that would not pose any particular difficulty. Let us imagine we want to analyze relative diffusion of two tracers in a cloud, which itself is rotated and advected by a larger scale atmospheric currents. In this case, the initial displacement between the particles matters. Rotation of the cloud would lead to changes in the relative displacement even if there is no diffusion inside the cloud. By focusing specifically on the statistics of the absolute distance between tracers we circumvent this issue. Interestingly, although previously introduced at least in some works~\cite{ueda:2015}, the statistics of the relative distance has not been studied in detail before.

\section{The mean-squared relative distance of two random walkers}
\label{sec:MSRD}
\begin{figure*}
	\resizebox{0.94\hsize}{!}{\includegraphics*{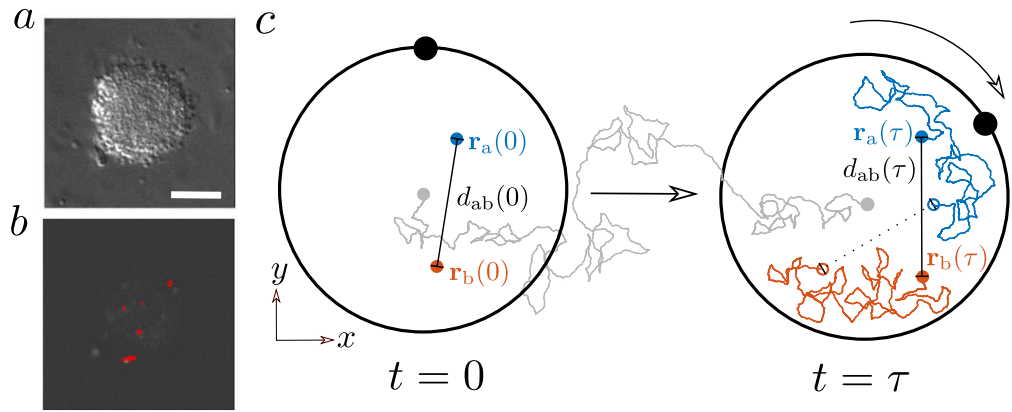}}
	\caption{Random motion of individual cells within aggregates. (a) Differential interference contrast microscopy image of a \textit{Neisseria gonorrhoeae} micrcolony consisting of $\sim$2000 individual cells. The scale bar is ${10\ \mu \mathrm{m}}$. (b) A small fraction of cells within a microcolony were fluorescently labeled. This allowed for tracking of their trajectories (red) with the help of a fluorescence channel of the microscope~\cite{ponisch20172}. (c) Sketch of a simplified two-dimensional aggregate. Two particles are initially separated by a distance $\dab(0)$. Up to a time $\tau$, they perform random motion in a circular domain, see solid lines illustrating the trajectories of particles.  However, the aggregate itself rotates (as marked by the black dot on the boundary of the domain) and its center of mass experiences some random motion (grey line). To quantify the diffusivity of particles we follow their absolute relative distance as a function of time $\dab(t)$, which is independent of the motion of the domain.}\label{fig1}
\end{figure*}

As an example, let us define the MSRD of two cells within a bacterial colony. Two cells, $a$ and $b$, at positions $\ra(t)$ and $\rb(t)$ have the absolute distance
\begin{equation}\label{eq:scalardistance}
\dab(t)= | \ra(t) - \rb(t) |.
\end{equation}
This quantity is independent of any translational or rotational motion of the cell aggregate. We define the MSRD, denoted $\MSRD(t)$, as the squared mean of the change of this distance with time $t$:
\begin{equation}\label{eq:MSRDdef}
\MSRD \left( t \right) = \langle \left( \dab(t) - \dab(0) \right)^2 \rangle.
\end{equation}
Here, we average over an ensemble of different realizations of $\dab(t)$ while keeping $\dab(0)$ fixed.
An illustration of the quantity $\MSRD(t)$ is shown in figure~\ref{fig1}c. For simplicity, but also in agreement with experimental measurements which are often taking place in a single focal plane of a microscope, we will consider a two-dimensional scenario. The generalization to higher dimensions is straightforward.

From figure~\ref{fig1}c it is easy to see that the MSRD is probably the only measurable quantity similar to the standard MSD but has the advantage of being insensitive to the motion of the cell aggregate as a whole. Our goal is to relate the behavior of $\MSRD(t)$ to the diffusivity of individual cells. 

\begin{figure*}\label{fig2}
	\resizebox{0.97\hsize}{!}{\includegraphics*{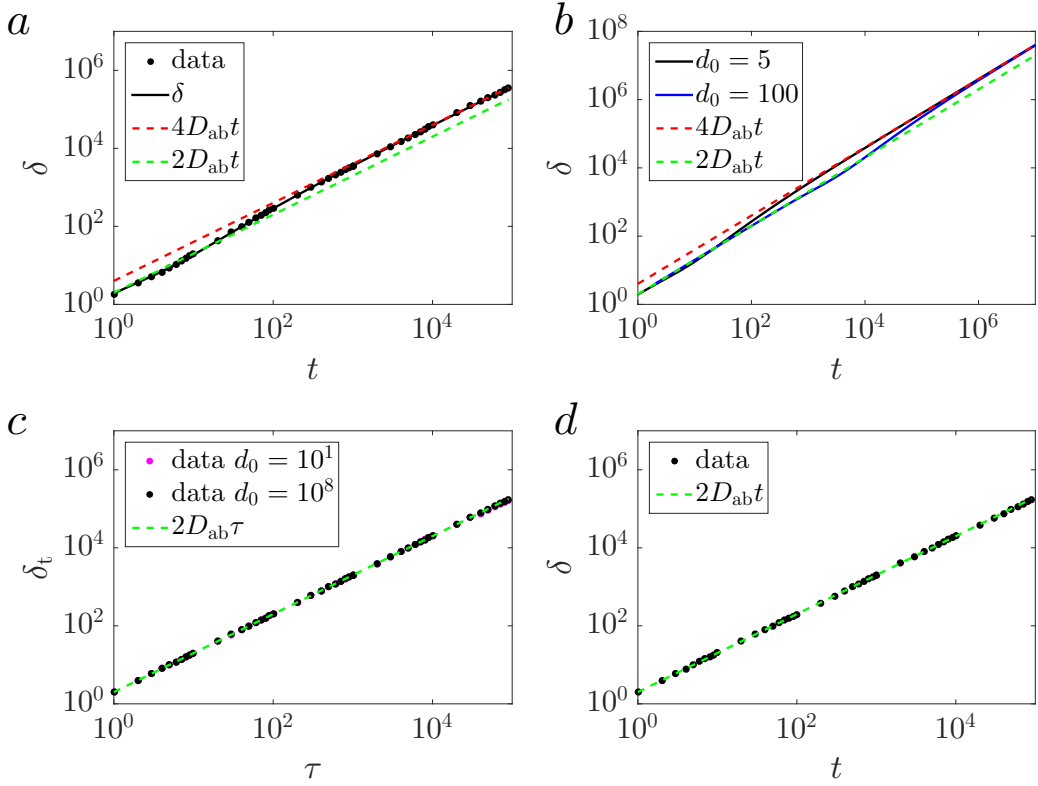}}
	\caption{Mean-squared relative distance for the random motion of two tracers in an unbounded domain. Here, ${\DAB=\DA+\DB=1}$ is the sum of the diffusion coefficients of the individual trajectories. (a) Ensemble-averaged MSRD for $d_0=4$ (dots). For small times $t$ the MSRD follows $2 \DAB t$ dependence and for large times it follows $4 \DAB t$ (dashed lines). Equation~\ref{eq:MSRDensemble} predicts the behavior of the MSRD for all times (solid line). (b) Ensemble-averaged MSRD for different initial conditions ${d_0=5,100}$, as given by equation~\ref{eq:MSRDensemble}. The transition between the two scaling regimes shift to later times for larger initial distance $d_0$. (c) Time-averaged MSRD for two particles with different initial distances ${d_0=10,10^8}$ (dots). For both initial distances the MSRD follows the function ${2 \DAB \tau}$ (dashed line). (d) Ensemble-averaged MSRD for randomly picked initial values $d_0$ follows the same dependence as the time averaged result. 
	}
\end{figure*}

In order to study the behavior of the MSRD, we first simulated the trajectories of pairs of Gaussian random walks with diffusion coefficients ${D = \DA = \DB = 0.5}$ (given as a unitless quantity) and an initial distance $d_0$ in an unbounded domain (details of the simulations can be found in appendix~\ref{sec:simulationdetails}). By computing the scalar distance of the two trajectories, defined in equation~\ref{eq:scalardistance}, we can compute the MSRD (see equation~\ref{eq:MSRDdef}) in the ensemble average sense (see figure~\ref{fig2}a). We observe that the MSRD exhibits two regimes which can be approximated as ${\MSRD ~\sim 4 D t }$ for small times $t$, and by ${\MSRD ~\sim 8 D t}$ for large times.
The transition point between the regimes depends on the initial distance $d_0$ with a later transition corresponding to a larger $d_0$ (see figure~\ref{fig2}b). 
We discuss the origin of the two regimes later in the text where the corresponding analytical expression for the MSRD is derived. In the transition region, the MSRD can be approximated by a power law with ${\MSRD \propto t^\alpha}$,\ ${\alpha > 1}$ (see figure~\ref{fig2}a). Of particular note, this transient behavior can be misinterpreted as a signature of superdiffusion if the time traces are not long enough to detect the second diffusive regime ~\cite{metzler2000}.

As an alternative to the ensemble-average, we computed the MSRD by time-averaging relative distances for a pair of very long trajectories. The time averaged MSRD $\MSRDtime$ is given by 
	\begin{equation}\label{eq:MSRDdef2}
	\MSRDtime \left( t \right) = \langle \left( \dab(t_0+ t) - \dab(t_0) \right)^2 \rangle_{t_0}.
	\end{equation}
	Here, for every lag time $t$, we average the relative distance over all starting points $t_0$ along the trajectory. Interestingly, we observe in our simulations that the MSRD follows a single scaling ${\MSRDtime \sim 4 D t}$ and is independent of the initial distance $d_0$ (see figure~\ref{fig2}c).

Often, the time-averaging is applied for the estimation of the diffusion coefficient in data where the statistics are not strong enough to deliver a reliable ensemble average. Examples for such cases are usually experiments in which a high effort is required to measure the trajectory of a single tracer, such as the study of the motion of individual tracers within cells or single cells themselves ~\cite{munder2016ph, tolic2004anomalous, caspi2000, goulian:2000, golding2006}. However, as became apparent recently, care should be taken when interpreting the time-averaged data \cite{metzler2014}.

Differences between time-averaged and ensemble-averaged quantities appear quite frequently, for example, if the tracers of the ensemble are in different dynamic states~\cite{qian:1991, he:2008}, or if the diffusion coefficient is not spatially homogeneous~\cite{cherstvy2013}. Additionally, time-averages and ensemble-averages can differ for the case of the so called weak ergodicity breaking which can be linked to power-law distributed waiting times present in a system of interest \cite{bel2005}. Examples of such systems are subdiffusive continuous time random walks and L\'{e}vy walks ~\cite{bel2005, burov2011, zaburdaev2015}, see also a recent review \cite{metzler2014}.

In our case, the difference of the ensemble-averaged and the time-averaged MSRD stems from the difference in the initial conditions of the random walks. While we picked the same initial condition $d_0$ for the computation of the ensemble-averaged MSRD, it follows from the definition of the time-averaged MSRD (see equation~\ref{eq:MSRDdef2}) that the initial condition is constantly changing. This idea is further supported by performing an additional averaging over the ensemble-averaged MSRD with respect to randomly chosen initial distances $d_0$. In this case, we observe that the MSRD shows the same behavior as the time-averaged one (see figure~\ref{fig2}d).

Next, we analytically calculate the ensemble-averaged and time-averaged MSRD. This allows us to explain the origin of the two regimes of the ensemble-averaged MSRD and the difference between ensemble-averaged and time-averaged MSRDs.

\section{Ensemble- and time-averaged mean squared relative distance}
\label{sec:ensembleandtime}
In order to calculate the ensemble-averaged MSRD, we first reduce the motion of two particles (denoted by $\mathrm{i}=a,b$) to the effective motion of a single particle.

The probability density functions of each particle position, $\ra$ and $\rb$, defined in cartesian coordinates $(x,y)$, are given by a Gaussian distribution
\begin{equation}
\pI (x, y, t) = \frac{1}{4\pi \Di t} \exp \left( - \frac{(x-\xzeroi)^2 + (y-\yzeroi)^2}{4 \Di t} \right),
\end{equation}
with the diffusion coefficient $\Di$ and the initial position $(\xzeroi, \yzeroi)$. The probability density function of the distance vector ${\dabvec(t)=\ra(t) - \rb(t)}$ of these two particles, starting with an initial distance ${d_0 = | \dabvec(0) |}$  in $y-$direction, is then given by 
\begin{equation}\label{eq:pAB}
\pAB (x,y,t) = \frac{1}{4\pi \DAB t} \exp \left[ - \frac{x^2 + (y - d_0)^2}{4 \DAB t} \right].
\end{equation}
This corresponds to the probability density function of a Gaussian random walker with a starting position $(0,d_0)$ and a diffusion coefficient of ${\DAB=\DA+\DB}$.

For the particular case of ${d_0=0}$, the probability density function of the scalar distance $\dab$ of a Gaussian random walk is given by the Rayleigh distribution~\cite{beckmann1962, siddiqui1962},
\begin{equation}\label{eq:Rayleigh}
\pRAYLEIGH (\dab,t) = \frac{\dab}{2 \DAB t} \exp \left( -\frac{\dab^2}{4 \DAB t}\right).
\end{equation}
For an arbitrary initial distance $d_0$, the probability density function of the scalar distance $\dab$ is given by the Rice distribution~\cite{wax1954},
\begin{equation}\label{eq:Rice}
\pRICE (\dab,t) = \frac{\dab}{2 \DAB t} \exp \left( -\frac{\dab^2 + d_0^2}{4 \DAB t}\right) I_0 \left( \frac{\dab d_0}{2 \DAB t}\right),
\end{equation}
where $I_x$ is the modified Bessel function of the first kind~\cite{abramowitz1966handbook}. The fact that the Rice distribution characterizes the distribution of the relative distance of two normally distributed particles is well known (see~\cite{yuce:l2013, buchin2012, buchin2015}).
This equation is frequently used in radar and sonar signal processing. For completeness, we have included the derivation of the Rice Distribution in Appendix~\ref{sec:ricederivation}.

In order to compute the MSRD, we calculate the first two moments of this probability density function:
\small
\begin{eqnarray}
\langle \dab (t) \rangle && = \frac{\sqrt{\pi}}{4 \sqrt{\DAB t}} \exp \left[ - \frac{d_0^2}{8 \DAB t} \right] \nonumber \\
&& \times \left[ (d_0^2 + 4 \DAB t)  I_0\left(\frac{d_0^2}{8 \DAB t}\right) + d_0^2 I_1\left(\frac{d_0^2}{8 \DAB t}\right)\right]  \\
\langle \dab^2 (t) \rangle && = d_0^2 + 4 \DAB t.\label{eq:secondmom}
\end{eqnarray}
\normalsize
Thus, the ensemble-averaged MSRD is given by
\begin{eqnarray}\label{eq:MSRDensemble}
\MSRD && = \langle \left[\dab(t) - \dab(0) \right]^2 \rangle \nonumber \\ && =\langle \dab^2(t) \rangle - 2 d_0 \langle \dab(t) \rangle + d_0^2  
\end{eqnarray}
where $\dab(0) = d_0$. Alternatively, one can also compute the MSRD by taking the probability density function of the distance vector between the two particles and computing the mean value of the mean squared scalar distance:
\small
\begin{eqnarray}\label{eq:MSRD1}
\MSRD &&= \langle ( \dab (t)  -  \dab (0) )^2 \rangle_{\pAB} \nonumber \\ &&= \iint  \mathrm{d}x \mathrm{d}y\ \pAB (x,y,t) (\dab (t)  -  \dab (0))^2. 
\end{eqnarray}
\normalsize
In both cases we arrive at the same result. The calculated ensemble-averaged MSRD (equation~\ref{eq:MSRDensemble}) reproduces the results of the numerical simulations (see figure~\ref{fig2}a).
We can approximate
\begin{equation}
\MSRD(t) \sim
\begin{cases} 
4 \DAB t &\mbox{if }{d_0^2 \ll \DAB t} \\
2 \DAB t & \mbox{if } {d_0^2 \gg \DAB t} 
\end{cases},
\end{equation}
thus the MSRD agrees with the observed limits  (see figure~\ref{fig2}a). The two regimes of diffusion exist due to the effect of the initial condition $d_0$. For small times, $d_0$ is much larger than the relative displacement due to tracers' diffusion $\Delta \mathbf{d}_{\text{ab}}$, ${d_0^2 \gg \Delta \mathbf{d}_{\text{ab}} \sim \DAB t}$. By expanding Eq.(\ref{eq:MSRDensemble}) in this limit, we can show that the change of the distance $\dab$ for a small displacement $\Delta \mathbf{d}_{\text{ab}}$ is approximated by the projection of $\Delta \mathbf{d}_{\text{ab}}$ in the direction of $\ra - \rb$, mimicking a one-dimensional random motion and explaining why the MSRD follows ${\MSRD \sim 2\DAB t }$ scaling.  Later, when the the limit ${d_0^2 \ll \DAB t}$ is fulfilled, the distance $d_0$ can be neglected and the distribution of displacements can be approximated by the Rayleigh distribution (see equation~\ref{eq:Rayleigh}). In this case, the motion is fully two-dimensional and we get ${\langle \dab^2(t) \rangle_{\pRAYLEIGH} \sim 4\DAB t}$. While $d_0$ does not affect the scaling behavior for large or small values of the time $t$ (compared to ${d_0^2 \slash \DAB}$), it does determine the transition time between these two limits, as can be seen in figure~\ref{fig2}b.

In order to compute the time-averaged MSRD, we average the time-dependent probability density function of the distances (see equation~\ref{eq:Rice}) over a time interval $[0,T]$ and call the resulting probability density function $\pRICEsum$ (see Appendix~\ref{sec:msrdtderivation} for the derivation). Then, we compute the mean value of the ensemble-averaged MSRD for this distribution. The resulting time-averaged MSRD is given by
\begin{equation}
\MSRDtime (\dt) = 2 \DAB \dt, 
\end{equation}
(see Appendix~\ref{sec:msrdtderivation} for its derivation). This result agrees with the behavior observed in simulations (see figure~\ref{fig2}c) and is independent of the initial distance $d_0$. We can relate this result to the ensemble-averaged MSRD. In the calculation of the running time average, and in the limit of the trajectory length going to infinity, the initial distances between tracers entering averaging also grow infinitely with time. Thus, in the corresponding ensemble-average picture we would be operating in the regime, where the diffusive displacement is much smaller than the initial distance. Hence the time-averaged MSRD has the same asymptotic as the first scaling regime of the ensemble-averaged MSRD.

Until now, we neglected the role of boundary conditions while studying the MSRD. In most cases, the cells will move within aggregates that are spatially confined (see figure~\ref{fig1}a and \ref{fig1}b). In the next section we consider the effects of a finite domain size.

\section{Effects of the finite domain size}
\label{sec:radialboundary}

Often, tracers move within domains of finite size, for example individual molecules inside single cells~\cite{munder2016ph, golding2006, guo2014} or individual cells within cell aggregates~\cite{ponisch20172, wessel2015}.
\begin{figure*}
	\resizebox{0.97\hsize}{!}{\includegraphics{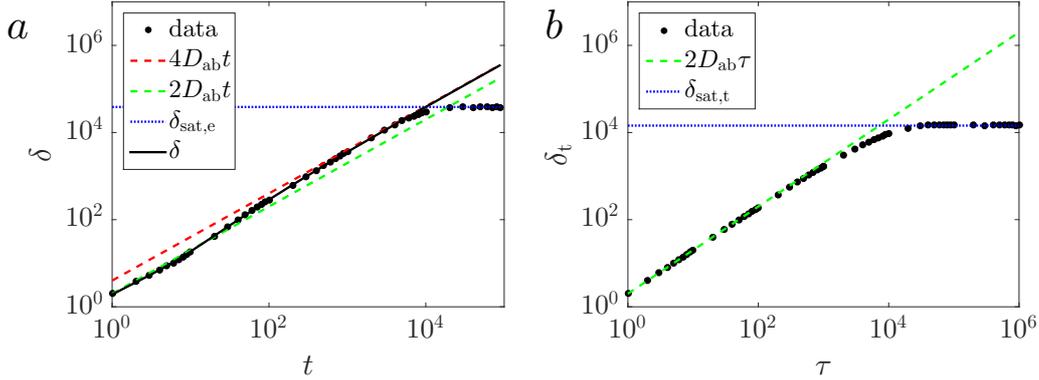}}
	\caption{Mean-squared relative distance for the random motion of two tracers moving within a circle. (a) Ensemble-averaged MSRD (dots) following equation~\ref{eq:MSRD1} (solid line) until the saturation MSRD $\MSRDsature$ (dotted line). (b) Time-averaged MSRD (dots) with the scaling ${2 \DAB \tau}$ (dashed line) and the saturation MSRD $\MSRDsaturt$ (dotted line).}
	\label{fig:figure3}
\end{figure*}
To account for such boundary effects, we simulated the motion of the two Gaussian random walkers within a circle (see figure~\ref{fig1}c) with reflective boundaries (details of the simulations are given in Appendix~\ref{sec:simulationdetails}). As might be expected, the behavior of the ensemble-averaged and time-averaged MSRD starts to be affected by the boundary when the displacement becomes comparable to the radius $R$ of the circle (see figure~\ref{fig:figure3}). For longer times, the MSRD saturates towards the values $\MSRDsature$ for the ensemble average and $\MSRDsaturt$ for the time average. The values of $\MSRDsature$ and $\MSRDsaturt$ can be estimated analytically (see Appendix~\ref{sec:additionalboundary}). The saturation values depend on the initial positions of the two particles for the ensemble-average and do not depend on the initial condition for the time average. The transition region towards saturation might be interpreted as a signature of subdiffusion. While subdiffusion might indeed occur in cells as a result of tracer particles being trapped in local environments, it is important to discriminate such behaviors from the effects of the domain (cell) boundary. In that respect, the analytical results of the MSRD saturation values $\MSRDsature$ and $\MSRDsaturt$ can provide an estimate on when to expect the influence of boundary effects.

\section{Conclusions}\label{sec:summary}
In this paper, we presented a tool to measure the diffusion coefficients of individual tracers within mobile domains. To mitigate the effect of domain movement, we suggest looking at the relative distance between pairs of tracer particles. Therefore, it is required to track the positions of at least two tracers simultaneously. From these data, one can measure the mean-squared relative distance of a pair of particles. The MSRD enables us to quantify the sum of the diffusion coefficients of motile cells within biological aggregates (for example bacterial microcolonies, cell spheroids or tissues), independent of translations and rotations of the agglomerates. Under the assumption of identical diffusivities of cells, this can be directly translated into the characteristics of the individual cells. 

In order to compute not just the sum of the diffusion coefficients, but their values, it is necessary to track not just two but three cells ($a,b,c$) simultaneously. By computing the three sums ${\DA+\DB}$, ${\DA+\DC}$ and ${\DB+\DC}$ it is possible to estimate the individual diffusion coefficients $\DA, \DB$ and $\DC$ of the individual cells. The method based on MSRD measurements can, with some limitations, be used for non-uniform diffusivities, where, for example, the diffusion constant is a function of the distance from the center of the aggregate~\cite{ponisch20172}. 

In this manuscript, we have shown that even in the simplest case of normal diffusion, analysis of the MSRD can exhibit some non-trivial behavior. The apparent diffusion constant read out from the ensemble-averaged MSRD may differ by a factor of 2 or even look like a superdiffusion in a transient regime. There is also a factor 2 difference in the long time scaling of the ensemble and time-averaged values of diffusion constant. We linked these differences to the initial separation between the tracers. Moreover, the domain size may lead to the saturation of the measured MSRD as a function of time. Our analytical results provide the guidelines for how the diffusivity of particles can be reliably extracted from the tracking data. This approach is viable for generalizing to anomalous (subdiffusion) and heterogeneous diffusion, which are both frequently encountered in biological settings~\cite{tolic2004anomalous, banks2005, bronstein2009}. There are differences between the time-averages and ensemble averages which contrasts the case of normal diffusion.  These differences are rooted in the underlying transport mechanisms, more robust and thus might be diagnostically relevant.
\appendix

\section{Simulation Details}\label{sec:simulationdetails}
In order to compute the ensemble-averaged MSRD, we simulated 2000 pairs of random walks in two dimensions. The walkers possessed a normally distributed step length and a constant step time, corresponding to a diffusion coefficient $D = 0.5$. In figure~\ref{fig2}d we studied the ensemble-averaged MSRD for random initial conditions. Therefore, the start points of the random walks where chosen independently of each other such that they were randomly distributed within a circle of radius $R=1000$.

In order to compute the time-averaged MSRD, we simulated two trajectories consisting of $10^7$ individual steps, with similar properties as the trajectories simulated for the ensemble-average. 

To consider the effect of boundaries, we included reflective boundary conditions relative to a circle of radius $R=200$. The initial positions of the random walks were chosen such that the first walker starts at the center of the circle and the second one starts at a distance $d_0 = 4$ from the first one.

\section{Derivation of the Rice distribution}\label{sec:ricederivation}
The Rice distribution can be derived by considering the probability density function of the scalar distance of a random walk from the origin of the coordinate system, given in equation~\ref{eq:pAB}. By computing the distribution of the distances $\dab$ and transforming the integral to polar coordinates, one can compute the distribution, which is defined to be the Rice distribution:
\begin{align}\label{eq:Rice2}
\pRICE (\dab,t) & = \iint \mathrm{d}x \mathrm{d}y\ \pAB(x,y,t) \delta (\dab - \sqrt{x^2+y^2}) \nonumber \\
& =  \iint \mathrm{d}R \mathrm{d}\phi\ R \pAB(R\cos\phi,R\sin\phi,t) \delta (\dab - R) \nonumber \\
& = \frac{\dab}{2 \DAB t} \exp \left[ -\frac{\dab^2 + d_0^2}{4 \DAB t}\right] I_0 \left( \frac{\dab d_0}{2 \DAB t}\right).
\end{align}
Here, $I_0$ is the modified Bessel function of the first kind~\cite{abramowitz1966handbook}.
\section{Time-averaged MSRD}\label{sec:msrdtderivation}
The probability density function of the time-dependent scalar distance $d(t)$ of two particles, performing a Gaussian random walk, is given by the Rice distribution (see equation~\ref{eq:Rice}). 

In order to compute the time averaged mean squared displacements of $d(t)$, we take this time-dependent distribution of distances and compute its mean over a time interval ${t \in [0, T]}$, given by 
\begin{equation}
\pRICEsum =  \lim\limits_{T \rightarrow \infty } \frac{\int_{0}^{T} \mathrm{d}t\ \pGRAY  \left( d, t\right) }{T}.
\end{equation}
The resulting equation, called $\pRICEsum$, represents the probability density function of having two particles with a distance $d$ at some point in the given time interval. ${t \in [0, T]}$

By performing the substitution $u = t \slash T$, this equation takes the form
\begin{equation}
\pRICEsum =  \lim\limits_{T \rightarrow \infty } \int_{0}^{1} \mathrm{d}u\ \frac{\dab}{2 \DAB u T} \exp \left[ -\frac{\dab^2 + d_0^2}{4 \DAB u T}\right] I_0 \left( \frac{\dab d_0}{2 \DAB u T}\right).
\end{equation}
The equation under the integral,
\begin{equation}
w(T)=\frac{\dab}{2 \DAB u T} \exp \left[ -\frac{\dab^2 + d_0^2}{4 \DAB u T}\right] I_0 \left( \frac{\dab d_0}{2 \DAB u T}\right),
\end{equation} 
is uniformly convergent~\cite{bronshtein2013handbook}, so that we can exchange the limit and the integral. For ${T \rightarrow \infty}$ the function within the integral converges
\begin{equation}
\lim\limits_{T \rightarrow \infty }\ w(T) = 0.
\end{equation}
This tells us that for an infinitely long time interval ${t \in [0, \infty]}$ all distances $d$ are equally likely and that there are no memory effects of the initial distance $d_0$. 

In the next step we compute the mean of the MSRD $\MSRD (d, \dt)$ for such an uniform distribution in the time interval ${t \in [0, g]}$ and then compute the limit ${g \rightarrow \infty}$. This is given by
\begin{equation}
\MSRDtime(\dt) = \lim\limits_{g \rightarrow \infty} \frac {\int_{0}^{g} \mathrm{d}d\  \MSRD (d, \dt)}{g } = 2 \DAB \dt.
\end{equation}
Again, we used a substitution of the form $u=d \slash g$ to simplify the integral. This is the time-averaged MSRD.

\section{Additional calculations for radial boundary conditions}\label{sec:additionalboundary}
For large times, the positions of two particles ($i=a,b$) within a circle of radius $R$ are uncorrelated and homogeneously distributed
\begin{equation}
\pCIRC (\ri)=\frac{1}{\pi R^2} \begin{cases} 1, &\ri \leq R \\
0, &\ri > R \end{cases}.
\end{equation}
The probability density function for two independent particles is then given by
\begin{equation}
\pCIRCab (\ra, \rb) = \pCIRC (\ra) \pCIRC (\rb).
\end{equation}
In order to compute the ensemble averaged saturation value of the MSRD, $\MSRDsature$, we define the distance of the two particles in polar coordinates ($R_1,\phi_1$) and ($R_2, \phi_2$)
\begin{align}
\dab =& | \ra - \rb | \nonumber \\ =& \sqrt{\left(R_1 \cos \phi_1 - R_2 \cos \phi_2\right)^2 + \left(R_1 \sin \phi_1 - R_2 \sin \phi_2\right)^2},
\end{align}
so that
\begin{align}
\MSRDsature(R,d_0)& = \langle  \left(\dab (t) - \dab (0) \right)^2 \rangle_{\pCIRCab} \nonumber \\
& = R^2 + d_0^2 - 2 d_0 f,
\end{align}
where
\small
\begin{align}
f(R) & = \frac{8}{\pi^2 R^4 } \times \nonumber  \int_{0}^{R} \mathrm{d}R_1 \int_{0}^{R} \mathrm{d}R_2\ \bigg[R_1 R_2 | R_1 - R_2|   \nonumber \\ & \times E \left( -\frac{4 R_1 R_2}{(R_1 - R_2)^2} \right) \bigg]  
\end{align}
\normalsize
and with $E(x)$ being the complete elliptic integral of $x$. Here, we define $\dab(0) = d_0$. This integral can be computed numerically.

The time-averaged saturation value of the MSRD is calculated by assuming that the initial positions of the particles are homogeneously distributed within the circle area $A$ and thus we integrate ${d_0 = |\razero - \rbzero|}$ over all positions of the two particles within the circle:
\begin{align}
\MSRDsaturt(R)=& \int_{A} \mathrm{d}\razero \int_A \mathrm{d}\rbzero \ \MSRDsature (d_0) \nonumber \\ =& 2 R^2 - 2 f^2.
\end{align}
Again, the resulting value of the MSRD $\MSRDsaturt$ can be computed numerically.
\section*{Acknowledgements}

We would like to thank Frank J\"ulicher and Nicolas Biais for fruitful discussions.
This work was supported by the Russian Science Foundation Grant No. 16-12-10496 (V.Z.).

\bibliography{bibliography}


\end{document}